\documentclass[pdflatex,sn-mathphys-num]{sn-jnl}%

\usepackage{graphicx}%
\usepackage{multirow}%
\usepackage{amsmath,amssymb,amsfonts}%
\usepackage{amsthm}%
\usepackage{mathrsfs}%
\usepackage[title]{appendix}%
\usepackage{xcolor}%
\usepackage{textcomp}%
\usepackage{manyfoot}%
\usepackage{booktabs}%
\usepackage{algorithm}%
\usepackage{algorithmicx}%
\usepackage{algpseudocode}%
\usepackage{listings}%
\usepackage{comment}
\usepackage{siunitx}

\begin{document}

\title[Diatoms modes
]{Diatom frustules as naturally occurring resonant microarchitectures: revealing vibrational eigenmodes across pennate and centric species}

\author*[1]{\fnm{Chiara} \sur{Gazzola}}\email{chiara.gazzola@polimi.it}

\author[2]{\fnm{Dame} \sur{Fall}}\email{dfall.iemn@gmail.com}

\author[3]{\fnm{Stefano} \sur{Stassi}}\email{stefano.stassi@polito.it}

\author[2]{\fnm{Marc} \sur{Duquennoy}}\email{Marc.Duquennoy@uphf.fr}

\author[4]{\fnm{Marc} \sur{Serra Garcia}}\email{m.serragarcia@amolf.nl}

\author*[5]{\fnm{Marco} \sur{Miniaci}}\email{marco.miniaci@gmail.com}

\affil*[1]{\orgdiv{Department of Civil and Environmental Engineering}, \orgname{Politecnico di Milano}, \orgaddress{\street{Piazza Leonardo da Vinci, 2}, \city{Milano}, \postcode{20133}, \country{Italy}}}

\affil[2]{\orgdiv{Univ. Polytechnique Hauts-de-France, CNRS, Univ. Lille, UMR 8520 - IEMN - Institut d’Électronique de Microélectronique et de Nanotechnologie, F-59313 Valenciennes, France}}

\affil[3]{\orgdiv{Department of Applied Science and Technology}, \orgname{Politecnico di Torino}, \orgaddress{\street{Corso Duca degli Abruzzi, 24}, \city{Torino}, \postcode{10129}, \country{Italy}}}

\affil[4]{\orgdiv{AMOLF Science Park 104, Amsterdam, The Netherlands}}

\affil[5]{\orgdiv{Department of Structural, Geotechnical and Building Engineering}, \orgname{Politecnico di Torino}, \orgaddress{\street{Corso Duca degli Abruzzi, 24}, \city{Torino}, \postcode{10129}, \country{Italy}}}

\abstract{Diatom frustules---the hierarchically structured, species-specific
silica exoskeletons of diatom microalgae---are among Nature's most
sophisticated examples of bottom-up self-assembly, exhibiting nanoscale
porosity, multifunctional mechanical and optical properties, and a
morphological diversity that spans nearly three orders of magnitude in size.
Despite growing interest in their optical and static mechanical properties,
the elastodynamic behaviour of frustules has remained largely unexplored.
Here we report the first experimental detection and spatially resolved
reconstruction of vibrational eigenmodes in diatom frustules, combining
laser Doppler vibrometry with morphology-faithful finite-element models
informed by scanning electron microscopy (SEM) and focused-ion-beam
scanning electron microscopy (FIB-SEM).
Two morphologically contrasting taxa were investigated as model systems:
the pennate diatom \textit{Rhaphoneis amphiceros} and the centric diatom
\textit{Stictodiscus californicus} var.\ \textit{nitida}, spanning the
two principal branches of diatom diversity.
Four eigenmodes were identified for \textit{R.\ amphiceros} in the
5.90--14.17\,MHz range and three for \textit{S.\ californicus} in the
9.96--17.62\,MHz range; the simulations yield a complete modal landscape
for each species, including modally split and nearly degenerate eigenmodes
arising from deviations from ideal symmetry, and show quantitative
agreement with the experimentally measured mode shapes and resonant
frequencies.
These results establish diatom frustules as a class of naturally occurring
resonant microarchitectures, and open new avenues for their integration
as bio-derived functional elements in nanomechanical and MEMS/NEMS
applications.}


\maketitle
\section{Introduction}\label{sec1}
Diatoms are photosynthetic unicellular microalgae that account for
approximately 20--25\,\% of global primary production and oxygen generation
\cite{armbrust2009life}, and whose silica exoskeletons---the
\textit{frustules}---represent a quintessential example of hierarchical
bottom-up self-assembly across multiple length scales
\cite{kroger2008diatoms,sumper2006learning}.
Each frustule is a species-specific, porous, multi-layered silica shell whose
characteristic dimensions span approximately 1 to \SI{500}{\micro\meter}
across the known flora \cite{round1990introduction, losic2009diatomaceous, de2017diatom}, with porosity regulated at the nano- to
microscale with high reproducibility.
Frustules are classified according to their macroscopic symmetry:
\textit{centric} species exhibit radial symmetry and are typically disc- or
cylinder-shaped, whereas \textit{pennate} species display bilateral symmetry
and adopt elongated, boat-like forms \cite{round1990introduction}.

These three-dimensional architectures surpass the structural complexity
attainable by state-of-the-art nanofabrication techniques, yet are synthesised
under ambient, environmentally benign conditions---a distinction that has
earned diatoms the designation of \textit{Nature's nanotechnologists}
\cite{bradbury2004nature, parkinson1999beyond, gordon2008glass,
losic2009diatomaceous, sumper2006learning, ragni2018multiple,
korsunsky2020siliceous}.
This bottom-up fabrication paradigm contrasts sharply with conventional
top-down lithographic approaches, which are inherently limited to planar
geometries and incur substantial energetic and economic costs
\cite{shalf2020future, gao2014environmental, lian2024quantifying}.
Bottom-up strategies can instead yield complex three-dimensional
nanoarchitectures \cite{rothemund2006folding, zhang2003fabrication,
whitesides1991molecular}, and the programmable morphological diversity of
diatom frustules---directly encoded in the genome and propagated through
self-replication \cite{armbrust2004genome,heintze2022molecular}---offers a scalable route to silica nanostructures with
tailored geometry \cite{parkinson1999beyond, jeffryes2011potential,
delalat2015targeted}.

Diatom frustules are multifunctional structures
\cite{de2017diatom, musenich2025revealing}.
Their architecture simultaneously confers mechanical stiffness against
predation-induced compression \cite{hamm2003architecture, aitken2016microstructure}
and enables transmittance of photosynthetically active radiation with
photonic-crystal-like spectral selectivity
\cite{fuhrmann2004diatoms, mcheik2018optical, romann_wavelength_2015,
detommasi2018uv}, while maintaining sufficient porosity to sustain efficient
solute exchange with the surrounding medium \cite{de2017diatom}.
This convergence of structural, optical, and transport properties---arising
from a single biogenic silica body shaped by hundreds of millions of years of
evolutionary selection---represents a design paradigm of broad interest to
materials science, photonics, and structural mechanics.
This potential has motivated their use as functional nanomaterials and as
biotemplates for the synthesis of siliceous and non-siliceous replicas
\cite{sandhage2002novel, zglobicka20193d, belegratis2014diatom,
cai2015biologically}, as well as inspiration for engineered materials
with improved mechanical \cite{musenich2025d} and optical
\cite{wang2026rational, xie2024diatom, li2019diatom} performance.

Despite this broad interest, the elastodynamic behaviour of diatom frustules
has received comparatively little attention.
From a basic-science perspective, the vibrational spectrum encodes the
interplay between frustule geometry, pore architecture, and material
properties, providing a sensitive probe of structure--function relationships
inaccessible to static measurements.
Resolving the elastodynamic response of frustules is thus a necessary step
toward uncovering the still-elusive link between morphogenesis, geometry, and
functional properties---a question whose significance is amplified by the
global ecological importance of diatoms among photosynthetic microorganisms
\cite{armbrust2009life}.
From an applied standpoint, diatom frustules share a close structural
resemblance to engineered micromechanical resonators---including MEMS
microphones \cite{je2013surface,cvjetinovic2023revealing}, coupled
nanomembrane resonators \cite{farah2026implementation} and optomechanical nanobeams \cite{leijssen2015strong}---suggesting their potential for direct integration
as bio-derived functional elements in nanotechnology platforms. The resonant frequencies of micrometric silica shells of this geometry
are expected to fall in the megahertz regime, placing them squarely
within the operational window of these same device classes.
Enabling their exploitation in MEMS/NEMS architectures, however, requires
a quantitative understanding of their vibrational behaviour that is
currently lacking.

Existing studies in this area are predominantly numerical.
\mbox{Gutiérrez et al.}\ \cite{gutierrez2017deformation} investigated the
mechanics of centric \textit{Coscinodiscus} sp.\ frustules through
finite-element modelling informed by SEM-derived morphology and experimentally
measured material properties, finding that the first ten deformation modes
closely match patterns observed in SEM images of naturally occurring
frustules---a result suggesting that these geometries may arise from
mechanically driven instabilities during morphogenesis.
Abdusatorov et al.\ \cite{abdusatorov2020fem} conducted a finite-element
parametric study of two morphologies---the centric \textit{Coscinodiscus}
sp.\ and the pennate \textit{Synedra acus}---demonstrating that their natural
frequencies lie in the MHz range and establishing approximate scaling relations
between eigenfrequencies, material properties, and geometric parameters.
The sole experimental characterisation of frustule vibrational eigenmodes was
subsequently reported by Cvjetinovic et al.\ \cite{cvjetinovic2023probing},
who combined finite-element modelling with atomic force microscopy to probe
the resonance behaviour of \textit{Coscinodiscus oculus-iridis} frustules.
Thermal-noise and piezoelectrically driven spectra acquired at multiple
locations on the frustule surface resolved several peaks in the 1--8\,MHz
range, which were tentatively assigned to specific eigenmodes through
numerical comparison.
While marking an essential first step, three fundamental limitations remain.
First, the experimental campaign yielded resonance spectra only: no spatially
resolved reconstruction of the associated mode shapes was performed.
Second, not all observed spectral features could be unambiguously mapped onto
simulated eigenmodes, partly because the numerical model adopted an idealised
circular geometry that neglects the actual frustule topology.
Third, structural heterogeneity and deviations from perfect rotational
symmetry---omnipresent in real frustules---can induce modal splitting and
pairs of nearly degenerate modes, effects that were not accounted for.

Here we report, to the best of our knowledge, the first experimental detection
and spatially resolved mapping of vibrational eigenmodes in diatom frustules
across a frequency range extending up to 20\,MHz.
We selected two morphologically contrasting taxa as model systems: the pennate
diatom \textit{Rhaphoneis amphiceros} (Ehrenberg) Ehrenberg, 1844
(class Bacillariophyceae) and the centric diatom \textit{Stictodiscus
californicus} var.\ \textit{nitida} Greville \& Sturt, 1887
(class Coscinodiscophyceae), thereby spanning the two principal branches of
diatom morphological diversity.
The structural geometry of each frustule was characterised by scanning
electron microscopy (SEM) and focused-ion-beam SEM (FIB-SEM), and directly
incorporated into finite-element models.
The resulting simulations yield a complete modal landscape for each species,
including modally split and nearly degenerate eigenmodes arising from symmetry
breaking.
Quantitative agreement between predicted and experimentally measured mode
shapes and resonant frequencies validates both the experimental methodology
and the modelling framework, and establishes diatom frustules as a class of
naturally occurring resonant microarchitectures whose vibrational properties
are accessible, reproducible, and potentially exploitable in bio-derived
nanomechanical applications.

\section{Results}\label{sec2}
\subsection*{Morphology acquisition and modelling}
Plan-view SEM imaging and FIB-SEM cross-sectional milling were performed
on single valves of \textit{Rhaphoneis amphiceros}
($\approx$110\,$\times$\,40\,\si{\micro\meter}) and
\textit{Stictodiscus californicus} var.\ \textit{nitida}
(diameter\,$\approx$\,49\,\si{\micro\meter}), isolated on standard
laboratory glass coverslips (Fig.\,1).
The two taxa present markedly different valve architectures.
The pennate valve of \textit{R.\ amphiceros} exhibits an elongated,
boat-shaped morphology whose face is perforated by a well-ordered areolar
array with locally square symmetry, the individual pores separated by a
continuous silica framework (Fig.\,1a,b).
The centric valve of \textit{S.\ californicus} var.\ \textit{nitida},
by contrast, presents a disc-shaped morphology with a circular valve face
perforated by areolae whose spatial arrangement reflects the underlying
radial symmetry of the taxon, with a more irregular pore distribution
than that observed in the pennate species (Fig.\,1d,e).
Despite these morphological differences, FIB-SEM cross-sectional imaging
reveals a suspended-plate structural motif common to both taxa: the
areolae extend through the full valve thickness, with no additional structural layers present (Fig.\,1c,f).

\begin{figure} 
	\centering
    \includegraphics[width=1\textwidth]{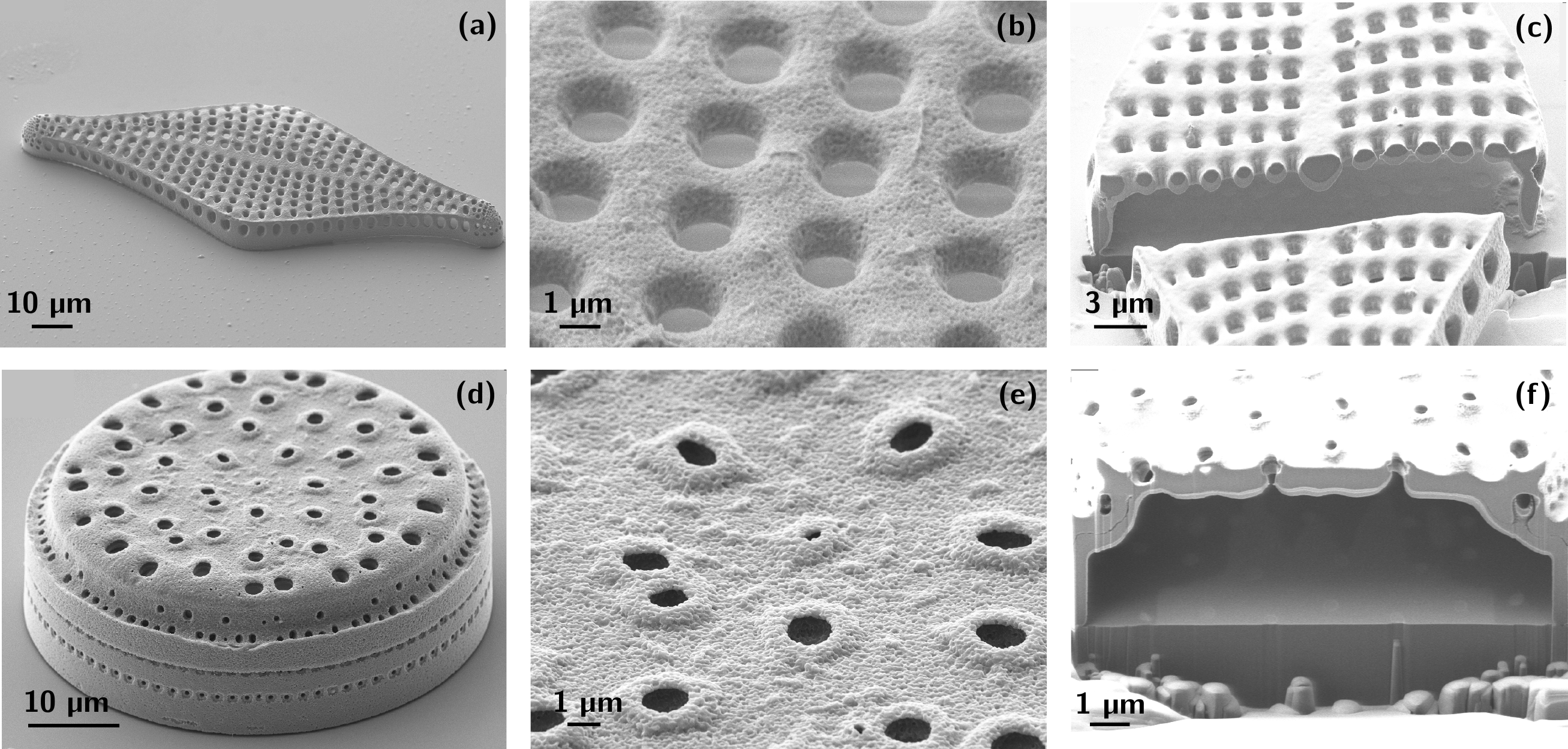}
    \caption{\textbf{Frustule morphology acquisition.}
SEM imaging characterises the in-plane pore architecture of each valve:
overview images of \textit{Rhaphoneis amphiceros} (\textbf{a}) and
\textit{Stictodiscus californicus} var.\ \textit{nitida} (\textbf{d})
illustrate the contrasting pennate and centric morphologies, while
high-magnification images (\textbf{b},\,\textbf{e}) resolve the valve pore
arrangement, revealing a locally square pore lattice in
\textit{R.\ amphiceros} and a more irregular pore distribution in
\textit{S.\ californicus}. FIB-SEM cross-sectional imaging probes the out-of-plane wall structure (\textbf{c},\,\textbf{f}), confirming that the pores extend through the full thickness of the valve and that both frustules consist of a single suspended silica plate with no additional structural layers.}
	\label{SEM_FIB-SEM}
\end{figure}
Since FIB-SEM sectioning is inherently destructive, the specimen used for
geometric characterisation cannot subsequently be tested vibrationally;
wall-thickness values are therefore representative of the population
rather than exact measurements of the specific frustule under test.
Since resonant frequencies of thin plate resonators scale as
$f \propto h/L^2$, where $h$ is the wall thickness and $L$ a
characteristic lateral dimension, intraspecific variability in wall
thickness propagates directly into the eigenfrequency predictions as
$\delta f/f \approx \delta h/h$, and is expected to be a source of
discrepancy between predicted and measured resonant frequencies.

Plan-view SEM images were binarised to extract the two-dimensional pore
topology of each valve and extruded to three-dimensional shell geometries
using the FIB-SEM-derived wall thickness, with the real pore arrangement
retained in full rather than homogenised, and the resulting geometries
meshed for finite-element modal analysis (Fig.2a).
The three-dimensional frustule wall architecture is approximated by a
planar extrusion of uniform or piecewise-uniform thickness.
The valve periphery was modelled as fully clamped.
Biogenic silica was modelled as a homogeneous, isotropic, linearly elastic
material with density $\rho = 2300$\,kg\,m$^{-3}$ and Poisson's ratio
$\nu = 0.2$.
For the Young's modulus, values reported in the literature for diatom
frustules span a wide range, as measurements reflect not only the
intrinsic properties of biogenic silica but also the specific pore
geometry, hierarchical organisation, and hydration state of each species
\cite{almqvist2001micromechanical, losic2007afm, cvjetinovic2023revealing,
moreno2015integrated}; since the pore geometry is explicitly resolved in
the present model, the adopted value represents the elastic modulus of
solid biogenic silica rather than an apparent modulus of the porous
frustule wall.
An upper bound of E = 72.2\,GPa, corresponding to the reported Young's modulus of amorphous silica \cite{pabst2013elastic}, was adopted as the starting value; this was then tuned to best match the first resonant frequency measured for both frustules, yielding E = 50\,GPa as the value used in the model.
\begin{figure} 
\centering
 \includegraphics[width=1\textwidth]{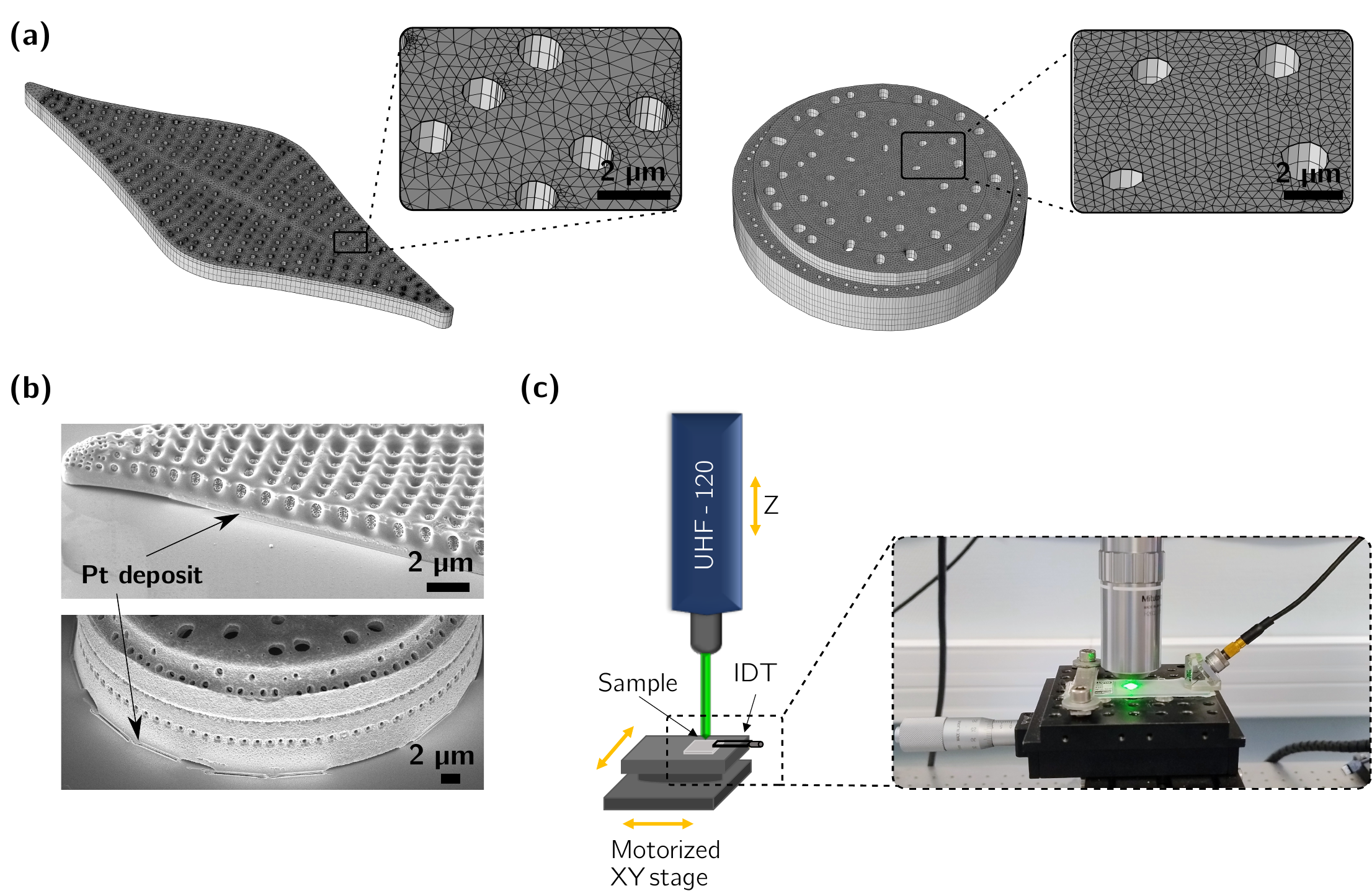} 
	\label{Raphoneis}
   \caption{\textbf{Finite-element models and experimental setup.}
(\textbf{a}) Specimen-specific finite-element meshes of \textit{Rhaphoneis
amphiceros} (left) and \textit{Stictodiscus californicus} var.\
\textit{nitida} (right), constructed by extruding binarised plan-view SEM
images to three-dimensional shell geometries using the FIB-SEM-derived
wall thickness; insets show details of the mesh at the pore scale,
illustrating the explicit representation of the real pore topology.
(\textbf{b}) FIB-SEM images of \textit{R.\ amphiceros} (top) and
\textit{S.\ californicus} (bottom) after focused-ion-beam-assisted
platinum deposition along the valve rim, realising the clamped boundary
condition imposed in the finite-element models and required for
vibrational testing.
(\textbf{c}) Schematic (left) and photograph (right) of the scanning
laser Doppler vibrometry setup.
 }
\end{figure}

\subsection*{Vibrational eigenmode mapping}
\begin{figure} 
	\centering
\includegraphics[width=1\textwidth]{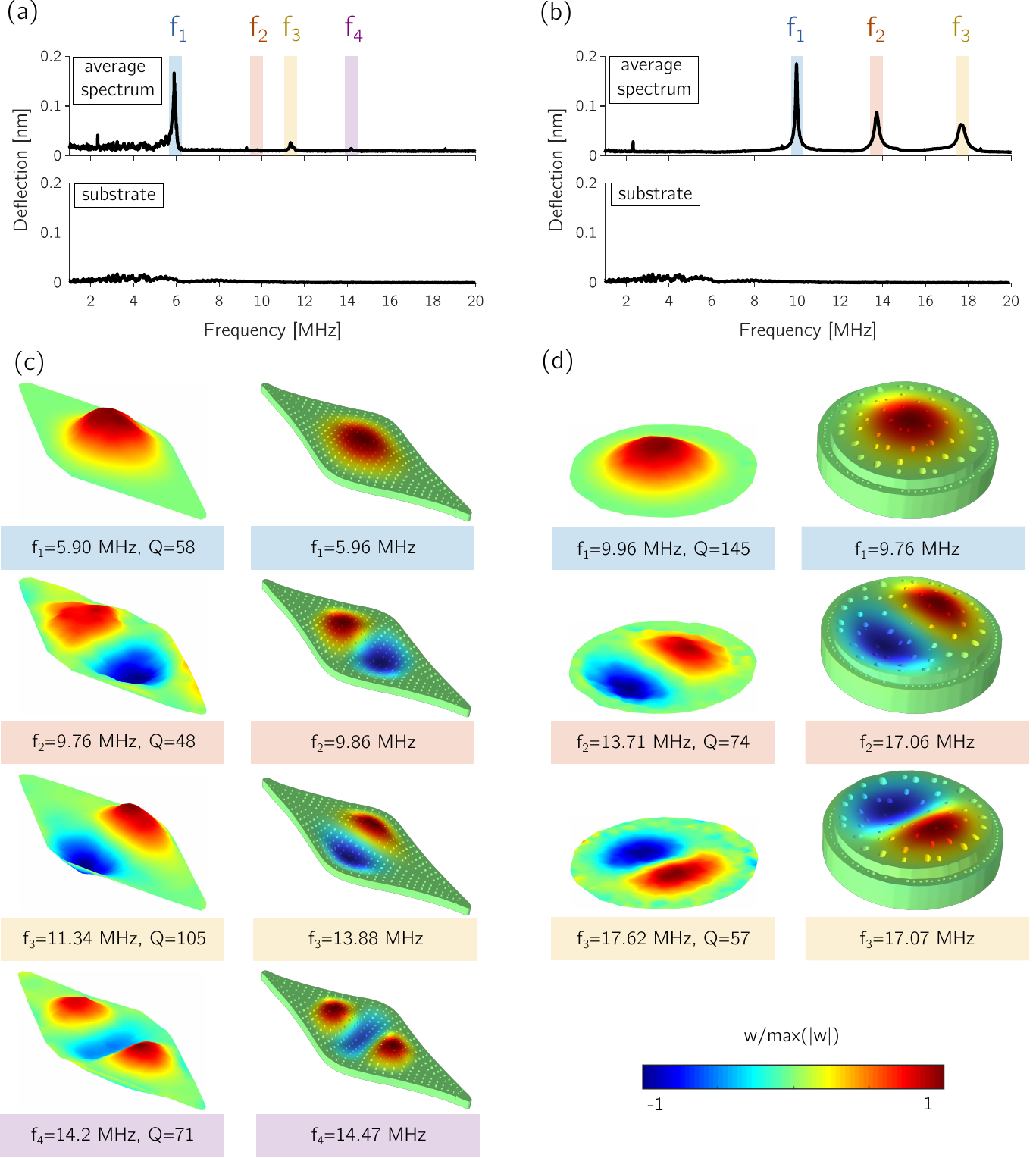} 
	\label{Raphoneis}
   \caption{\textbf{Experimental and numerical vibrational eigenmodes}. Deflection spectra measured by scanning laser Doppler vibrometry and obtained by averaging the out-of-plane response over all scanned points on the frustule surface, together with the substrate reference response, for \textit{Rhaphoneis amphiceros} (\textbf{a}) and \textit{Stictodiscus californicus} var.\ \textit{nitida} (\textbf{b}). Mechanical resonance peaks are highlighted and labelled $f_1$--$f_4$ and $f_1$--$f_3$, respectively. (\textbf{c},\textbf{d}) Experimentally measured spatially resolved mode shapes (left column) and corresponding finite-element simulations (right column) for the identified eigenmodes of \textit{R.\ amphiceros} (\textbf{c}) and \textit{S.\ californicus} (\textbf{d}). Experimental and simulated mode shapes are shown using the same normalised out-of-plane displacement scale, $w/\max(|w|)$, with colours ranging from $-1$ (blue) to $+1$ (red), corresponding to opposite phases of oscillation. Resonance frequencies and experimentally determined quality factors ($Q$) are reported for each mode, while the corresponding simulated eigenfrequencies are given for comparison.
 }
\end{figure}
A clamped boundary condition between frustule and substrate was realised
by securing individual valves to the glass coverslip through
focused-ion-beam-assisted platinum deposition at the valve rim
(Fig.\,2b).
Vibrational eigenmodes were then mapped by scanning laser Doppler
vibrometry over a spatial grid covering the full valve surface, with
acoustic excitation delivered by wedge ultrasonic transducers coupled to
the glass substrate beneath the sample; a photograph of the experimental
setup and a schematic of the measurement configuration are shown in
Fig.\,2c (see Methods for full acquisition parameters).
The laser beam was raster-scanned across the valve surface and a full
vibratory spectrum recorded at each grid point; spatially averaged
spectra were used to identify resonant frequencies, and mode shapes were
reconstructed by extracting the amplitude and phase of the out-of-plane
displacement at each grid point for every identified resonance.
A substrate reference spectrum, acquired under identical conditions on the
bare glass coverslip adjacent to the frustule, confirmed that all
identified spectral features originate from the frustule itself.

The spatially averaged LDV spectra acquired over the valve surfaces of
\textit{Rhaphoneis amphiceros} and \textit{Stictodiscus californicus}
var.\ \textit{nitida} reveal well-defined resonance peaks against a flat
noise floor, with no corresponding features in the substrate reference
spectra (Fig.\,3a,b).
Four resonant frequencies were identified for \textit{R.\ amphiceros}
in the 1--20\,MHz range ($f_1 = 5.90$\,MHz, $f_2 = 9.76$\,MHz,
$f_3 = 11.34$\,MHz, $f_4 = 14.17$\,MHz), and three for
\textit{S.\ californicus} ($f_1 = 9.96$\,MHz, $f_2 = 13.71$\,MHz,
$f_3 = 17.62$\,MHz).

The spatially resolved out-of-plane displacement maps reconstructed at
each resonant frequency (Fig.\,3c,d) reveal the mode shapes of both
frustules and are discussed below in conjunction with the corresponding
finite-element predictions.
For \textit{R.\ amphiceros}, the reconstructed mode shapes follow the
bilateral symmetry of the pennate valve.
The fundamental mode $f_1$ is a monopole mode characterised by a single
displacement antinode localised at the centre of the valve, with
displacements decaying almost symmetrically towards the clamped peripheral rim.
The second mode $f_2$ displays two antinodes distributed along the
longitudinal axis, consistent with the first longitudinal bending mode
of the elongated pennate geometry, while $f_3$ and $f_4$ exhibit
progressively more complex spatial patterns with three and four antinodes
respectively, reflecting higher-order flexural modes along the
longitudinal direction.
The frequency ratios $f_2/f_1 = 1.65$, $f_3/f_1 = 1.92$, and
$f_4/f_1 = 2.41$ deviate significantly from the integer multiples
expected for a uniform Euler--Bernoulli beam, reflecting the
influence of the non-uniform mass and stiffness distribution introduced
by the areolar pattern and the tapered valve morphology.
For \textit{S.\ californicus}, the fundamental mode $f_1$ is likewise
a monopole mode, presenting a single axisymmetric displacement antinode
centred on the valve, consistent with the $(0,1)$ mode of a clamped
circular plate.
The modes $f_2$ and $f_3$ correspond to a split pair of the $(1,1)$
diametral mode, each displaying two lobes of opposite phase separated
by a nodal diameter oriented orthogonally between the two; their
frequency ratios $f_2/f_1 = 1.38$ and $f_3/f_1 = 1.77$ depart from the value expected for an ideal clamped circular
plate ($f_{(1,1)}/f_{(0,1)} = 2.081$ \cite{schmid2016fundamentals}), reflecting the
influence of the real pore arrangement and geometric imperfections on
the modal structure.

In both specimens the fundamental monopole mode dominates the vibrational
response at the lowest frequency, and the fundamental frequency of
\textit{R.\ amphiceros} ($f_1 = 5.9$\,MHz) is lower than that of
\textit{S.\ californicus} ($f_1 = 9.96$\,MHz), consistent with the
larger in-plane dimensions and elongated geometry of the pennate valve.

The finite-element predictions are in good overall agreement with the
experimental results in both mode shape topology and resonant frequency.
For \textit{R.\ amphiceros}, the relative frequency error
$\Delta f/f_\mathrm{exp}$ is below 2\,\% for $f_1$, $f_2$, and $f_4$
(simulated frequencies: 6.0, 9.8, and 14.5\,MHz respectively), and the
simulated mode shapes are in excellent visual agreement with the
experimental displacement maps.
The larger discrepancy observed for $f_3$ (22\,\%, 13.9 vs 11.4\,MHz)
is consistent with the dipolar character of this mode, whose nodal line
passes close to the valve rim and whose frequency is therefore sensitive
to the local bending stiffness at the periphery---a region where the
frustule wall exhibits out-of-plane features not fully captured by the
piecewise-uniform extrusion model.
For \textit{S.\ californicus}, $f_1$ is predicted within 2\,\% of the
experimental value (9.76 vs 9.96\,MHz), while the two diametral modes
$f_2$ and $f_3$ show discrepancies of 24\,\% and 3\,\% respectively.
The significant mismatch for $f_2$ is similarly attributable to its
dipolar character: diametral modes are particularly sensitive to the
peripheral stiffness distribution, and FIB-SEM cross-sections of
\textit{S.\ californicus} reveal a clear out-of-plane asymmetry at the
valve rim---with structural features present on one face but absent on
the other---that is not reproduced in the finite-element model.
More generally, the central region of the frustule wall is more
homogeneous and well-captured by the extrusion geometry, whereas the
peripheral rim exhibits richer out-of-plane architecture whose influence
is most pronounced for modes with significant deformation at the boundary.
Notably, the simulated frequencies of $f_2$ and $f_3$ are nearly
degenerate (17.06 and 17.07\,MHz): despite the use of the real
specimen-specific pore arrangement and valve contour in the model, the
near-radial symmetry of \textit{S.\ californicus} results in two
diametral modes with almost identical eigenfrequencies.
Their experimentally observed frequency splitting ($\Delta f = 3.91$\,MHz)
thus reflects symmetry-breaking features of the real frustule---such as
thickness gradients and out-of-plane rim asymmetry---that are present in
the physical specimen but not fully captured by the piecewise-uniform
extrusion model.

Quality factors were extracted by fitting a Lorentzian lineshape to each
resonance peak in the spatially averaged spectrum, and span
$Q = 48$--$105$ for \textit{R.\ amphiceros} and $Q = 57$--$145$ for
\textit{S.\ californicus}, with the monopole mode consistently yielding
the highest $Q$ in both taxa ($Q = 58$ and $Q = 145$, respectively).

\section*{Discussion}
\begin{figure} 
	\centering    \includegraphics[width=1\textwidth]{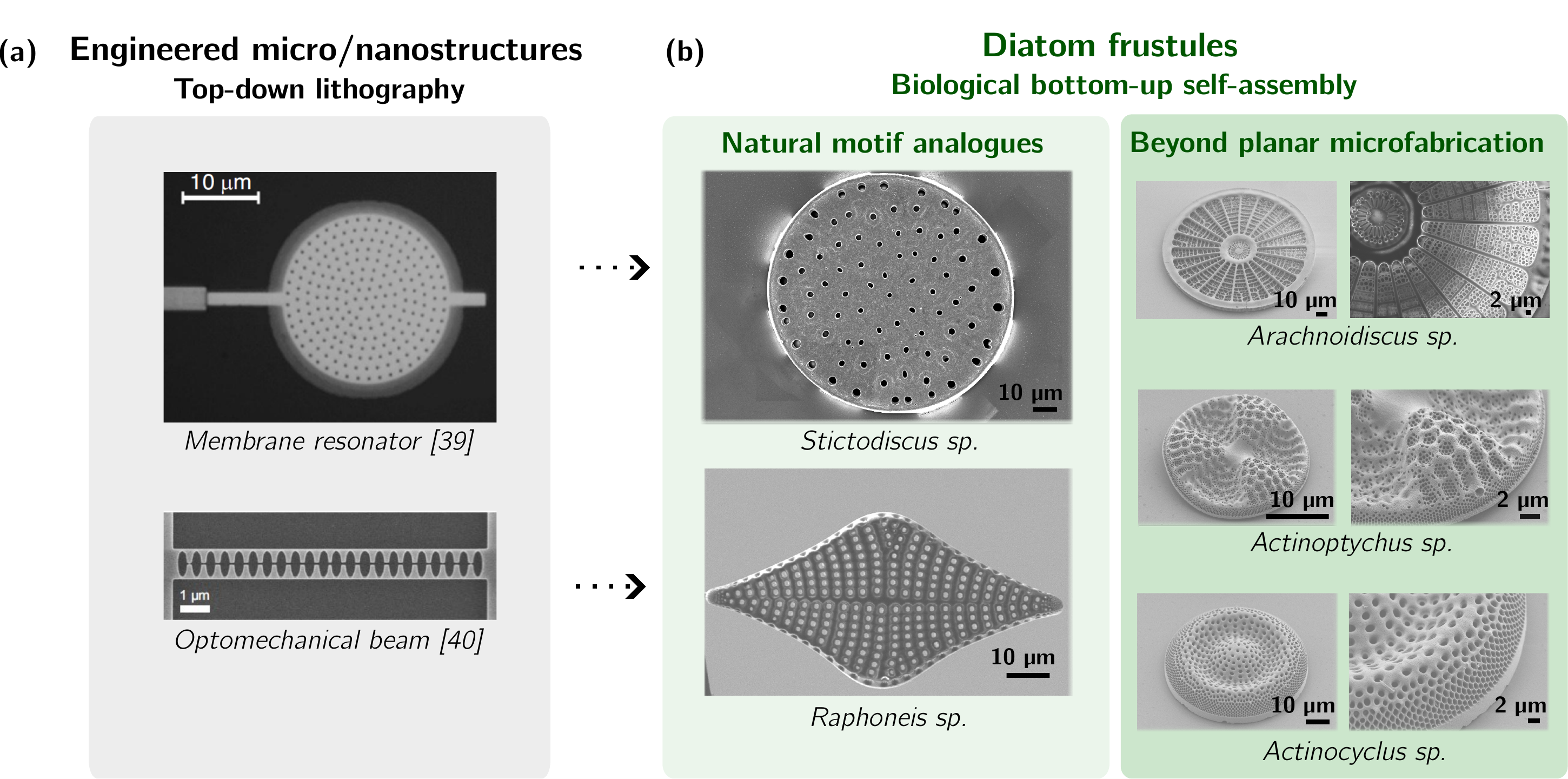} 
	\label{Raphoneis}
   \caption{\textbf{Diatom frustules as bio-derived analogues of engineered
microstructures.}
(a) representative engineered MEMS/NEMS devices fabricated by top-down
lithography, including a membrane resonator \cite{farah2026implementation}, and an optomechanical nanobeam \cite{leijssen2015strong}.
(b) Left: SEM images of the two frustule taxa investigated in this study,
\textit{Stictodiscus californicus} var.\ \textit{nitida} (top) and
\textit{Rhaphoneis amphiceros} (bottom), whose architectural motifs
closely parallel those of the membrane resonator and the
optomechanical beam, respectively.
Right: selected diatom species---\textit{Arachnoidiscus} sp.,
\textit{Actinoptychus} sp., and \textit{Actinocyclus} sp.---illustrating
the broader morphological diversity of the diatom flora, which encompasses
curved, hierarchically structured, and three-dimensional silica
architectures that lie beyond the geometric reach of standard planar
lithographic fabrication. Membrane resonator panel adapted under the terms of the CC-BY Creative Commons
    Attribution 4.0 International License
    (\url{http://creativecommons.org/licenses/by/4.0/}).~\cite{farah2026implementation}
    Copyright 2026, The Authors, Published by Springer Nature.
    Optomechanical beam panel adapted under the terms of the CC-BY Creative Commons
    Attribution 4.0 International License
    (\url{http://creativecommons.org/licenses/by/4.0/}).~\cite{leijssen2015strong}
    Copyright 2015, The Authors, Published by Springer Nature.
 }
\end{figure}
This work demonstrates simultaneous spectral and spatially resolved mapping
of vibrational eigenmodes in diatom frustules across two morphologically
contrasting taxa, with good quantitative agreement between experiment and
simulation.
Spatial mode shape reconstruction proves essential for a complete
interpretation of the vibrational response: the resonance spectrum alone
is insufficient to unambiguously assign spectral peaks to specific
eigenmodes, particularly in the presence of nearly degenerate mode pairs
arising from symmetry breaking induced by the real frustule topology.
This phenomenon---the lifting of modal degeneracy by geometric
imperfections---is well established in engineered nanostructures,
where fabrication-induced deviations from perfect rotational symmetry
produce pairs of spatially distinct modes with orthogonal nodal lines and
closely spaced resonant frequencies \cite{davidovikj2016visualizing}.
In diatom frustules, an analogous mechanism operates through the
species-specific porous architecture: the irregular pore boundaries,
thickness gradients, and rim geometry collectively break the ideal
symmetry of the valve and lift what would otherwise be degenerate mode
pairs.
The residual discrepancies between predicted and measured eigenfrequencies
are attributable to the geometric approximations inherent in the
reconstruction workflow.
In the present approach, the three-dimensional frustule wall is
represented by extruding a binarised plan-view mask to a uniform or
piecewise-uniform thickness derived from FIB-SEM cross-sections,
inevitably neglecting out-of-plane architectural features that influence
the local bending stiffness distribution.
A more faithful three-dimensional reconstruction would require either
continuous FIB-SEM serial sectioning \cite{xing2017characterization},
which provides nanometric resolution but remains destructive, or
nano X-ray computed tomography (nano-XCT) \cite{zglobicka20193d},
which is non-destructive but is currently limited to a spatial resolution
of approximately 50--100\,nm---potentially insufficient to resolve the
finest features of the frustule wall architecture.
The fundamental constraint therefore persists: achieving both
non-destructive imaging and nanometric resolution on the same individual
prior to vibrational testing remains an open instrumental challenge,
and represents a natural direction for future methodological development.

Comparison with engineered micro- and nanoelectromechanical devices reveals
a close structural and spectral correspondence, as illustrated in
Fig.\,4.
At the architectural level, \textit{Stictodiscus californicus} shares its
essential motif of a circular, peripherally anchored perforated membrane
with MEMS microphones \cite{je2013surface} and capacitive micromachined
ultrasonic transducers (CMUTs) \cite{butaud2020towards}, while
\textit{Rhaphoneis amphiceros}, with its elongated geometry and periodic
rib structure, finds close counterparts in optomechanical nanobeams
\cite{leijssen2015strong} and phononic crystal waveguides
\cite{zega2023defect}.
The analogy is particularly direct for the monopole mode of
\textit{S.\ californicus}: its axisymmetric $(0,1)$ breathing mode at
$f_1 = 9.96$\,MHz closely parallels the fundamental $(0,1)$ mode of
silicon nitride circular membrane resonators recently reported at
8.82\,MHz \cite{xu2024imaging} and 12.265\,MHz \cite{farah2026implementation},
which share the same clamped circular plate geometry and operate in the
same frequency window.
Yet whereas these engineered devices are fabricated through multi-step
top-down lithographic processes under tightly controlled conditions,
the frustule geometries emerge from a biological bottom-up self-assembly
process encoded in the genome and executed at ambient temperature and
pressure.
The resonant frequencies recovered experimentally---spanning 5.9 to
17.6\,MHz---fall squarely within the operational window of these same
device classes, placing diatom frustules in direct spectral overlap with
coupled membrane resonators explored for in-sensor computing
\cite{farah2026implementation}, CMUTs used in medical ultrasound imaging
\cite{butaud2020towards}, and cavity optomechanical devices
\cite{leijssen2015strong}.

A quantitative assessment of the reproducibility of the reported
eigenfrequencies across individuals of the same species nonetheless
remains to be established.
This question is non-trivial in diatoms: under asexual reproduction,
each cell division produces one daughter cell of reduced size, so that
a clonal culture maintained over multiple generations spans a continuous
distribution of frustule dimensions \cite{round1990introduction,
chepurnov2004}.
Since resonant frequencies of flexural plate modes scale with the ratio
of wall thickness to lateral dimension, intraspecific size variation
translates directly into a distribution of eigenfrequencies rather than
a single value---a feature that may itself be exploited, as frustules of
the same species but different dimensions would yield a natural spread
of resonant frequencies accessible within a single culture.
Characterising this distribution---by mapping the structural and
dimensional variability across a synchronised culture and correlating it
with the resulting spread in resonant frequencies---would clarify whether
intraspecific variability is sufficiently narrow to support the use of
diatom frustules as reproducible resonant components, or whether
size-selection strategies would be required to achieve the frequency precision demanded by specific
device contexts.

The present results, by establishing a validated experimental and
computational framework for frustule vibrational characterisation, provide
the methodological foundation for a systematic screening of the vast
morphological diversity of the diatom flora---spanning more than 100,000
described species \cite{de2017diatom} and encompassing curved,
hierarchically structured, and three-dimensional silica architectures
that lie beyond the geometric reach of standard planar lithographic
fabrication---with a view to identifying taxa whose resonant properties
are optimally matched to specific sensing, transduction, or computing
applications.

\section*{Methods}

\subsection*{Diatom specimens}

Diatom frustules were sourced from Stefano Barone (Diatom Shop, Italy).
Specimens were provided as dried valves individually mounted on standard
microscope glass slides, a preparation designed to preserve structural
integrity and facilitate direct observation of both valve-scale morphology
and pore-scale architecture.

\subsection*{Experimental vibrational characterisation}

Vibrational measurements were performed on the WAVESURF platform (IEMN,
Universit\'{e} Polytechnique Hauts-de-France, Valenciennes) using a
Polytec UHF-120 scanning laser Doppler vibrometer operating at a
wavelength of 532\,nm.
The instrument measures instantaneous out-of-plane surface velocity via
heterodyne interferometry; the laser beam was focused through a
$\times$100 objective, yielding a spot diameter of
0.88\,\si{\micro\meter}.
Spatial scanning was performed using a motorised $XY$ stage with a
positioning resolution of 500\,nm.
Photodiode signals were acquired using a LeCroy 725Zi-A oscilloscope
(2.5\,GHz, 40\,GS/s) and demodulated numerically via IQ demodulation
on a PC.
Acoustic excitation was delivered by wedge ultrasonic transducers coupled
to the glass substrate beneath the sample, which generate elastic waves
via ultrasonic refraction at the wedge--substrate interface, providing
broadband excitation across the frequency range of interest.
Transducers were driven by broadband linear temporal chirp signals of
1.5\,\si{\micro\second} duration, generated by a Tektronix AWG\,70002A
arbitrary waveform generator (2\,GHz, 16\,GS/s, 10-bit vertical
resolution) and amplified by an Amplifier Research 50W1000A RF power
amplifier (1--1000\,MHz, 50\,W), providing excitation voltages up to
50\,V$_{\mathrm{pp}}$.
Several transducers were evaluated for each specimen to optimise the
signal-to-noise ratio; the identified resonant frequencies were
consistent across transducer types, and the spectra presented in the
main text correspond to the configuration yielding the best measurement
quality.
The C5 wedge transducer was selected for \textit{Rhaphoneis amphiceros},
while the C10 wedge transducer---with IDT measurements used as
cross-validation---was selected for \textit{Stictodiscus californicus}.
For \textit{Rhaphoneis amphiceros}, measurements were performed with the
following parameters: sampling frequency 100\,MHz; acquisition duration
80\,\si{\micro\second}; frequency resolution 12.5\,kHz; bandpass filter
1--20\,MHz; 1521 FFT lines; spatial grid of 664 points with a step of
approximately 2\,\si{\micro\meter}.
For \textit{Stictodiscus californicus}, the acquisition parameters were:
sampling frequency 100\,MHz; acquisition duration 80\,\si{\micro\second};
frequency resolution 12.5\,kHz; bandpass filter 1--30\,MHz; 2321 FFT
lines; spatial grid of 1807 points with a step of approximately
1.2\,\si{\micro\meter}.

\subsection*{Finite-element modelling}
Specimen-specific finite-element models were constructed and solved in
COMSOL Multiphysics 6.4 (Structural Mechanics module).
Three-dimensional geometries were generated by binarising plan-view SEM
images to extract the real pore topology of each valve, extruding the
resulting two-dimensional mask along the normal direction by the
FIB-SEM-derived wall thickness, and meshing the volume with second-order
quadratic serendipity elements.
The mesh of \textit{Rhaphoneis amphiceros} comprised 100,414 wedge
elements, while that of \textit{Stictodiscus californicus} comprised
91,495 elements (25,811 tetrahedra, 280 pyramids, and 65,404 prisms).
Mesh convergence was verified by confirming that eigenfrequency
predictions varied by less than 2$\%$ upon further refinement.

\bibliography{BIBLIO.bib}
\end{document}